\documentclass[twocolumn,floatfix,prb,aps,showpacs]{revtex4-1}
\usepackage{graphicx,amsmath,amssymb}

\begin{document}

\title{Unpaired composite fermion, topological exciton, and zero mode}
\author{G. J. Sreejith,$^1$ A. W\'ojs,$^2$ and J. K. Jain$^{1,3}$}
\affiliation{$^1$Department of Physics, 104 Davey Lab, Pennsylvania State University, University Park PA, 16802}
\affiliation{$^2$Institute of Physics, Wroclaw University of Technology, 50-370 Wroclaw, Poland}
\affiliation{$^3$School of Physics, Korea Institute for Advanced Study, Seoul 130-722, Korea}

\date{\today}

\begin{abstract}
The paired state of composite fermions is expected to support two kinds of excitations: vortices and unpaired composite fermions.  We construct an explicit microscopic description of the unpaired composite fermions, which we demonstrate to be accurate for a 3-body model interaction, and, possibly, adiabatically connected to the Coulomb solution. This understanding reveals that an unpaired composite fermion carries with it a charge-neutral ``topological" exciton, which, in turn, helps provide microscopic insight into the origin of zero modes, fusion rules, and energetics.
\end{abstract}

\pacs{73.43.Cd, 71.10.Pm}
\maketitle

Should composite fermions (CFs) form a p-wave paired state, as has been proposed\cite{theory5p2,Greiter} for the mechanism of the fractional quantum Hall effect (FQHE) at 5/2, the analogy to superconductivity leads one to expect two kinds of excitations: vortices and unpaired CFs (UCFs). We construct in this Letter a microscopic description of the UCF\cite{comment1} by considering a state with an odd number ($N$) of composite fermions, which necessarily contains a composite fermion without a partner. The understanding of its structure reveals that the UCF carries with it what Hansson has termed a ``topological"-exciton,\cite{Hansson,Hansson09} and entails a number of remarkable and physically transparent properties for energetics, such as even-odd oscillations and zero modes, as well as fusion rules. Several of these have been found previously in the exact diagonalization studies of M\"oller, W\'ojs and Cooper,\cite{Moller} Bonderson, Feiguin and Nayak,\cite{Bonderson} and Lu, Das Sarma and Park,\cite{Hantao} and also in the conformal field theory treatment of the 5/2 state.\cite{theory5p2,Nayak,NayakRMP}

\begin{figure}
\includegraphics[width=0.40\textwidth]{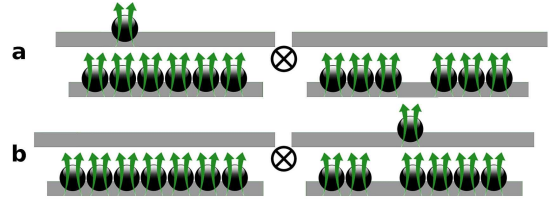}
\caption{(color online)
\label{schematic}
Schematic depiction of (a) the unpaired composite fermion with topological exciton; (b) ordinary exciton.  The composite fermions are shown as dots decorated with arrows, representing bound states of electrons and vortices. The left and right parts show the $\Lambda$ level diagrams for composite fermions in the two partitions. The single CF in the otherwise empty $\Lambda$ level is called a (CF-)quasiparticle and the missing CF a (CF-)quasihole, which, for the present case, have charge excess or deficiency of magnitude $e/4$ relative to the uniform ground state.
\vspace*{-6mm}
}
\label{fig1}
\end{figure}

We consider the following ansatz for the UCF state:
\begin{equation}
\Psi^{\rm UCF}={\cal A}\;\phi^{\rm CF}(\{z_j\}) \; \chi^{\rm CF}(\{w_k\}) \prod_{j=1}^{M+1} \prod_{k=1}^{M}(z_j-w_k)
\nonumber
\end{equation}
where the $N=2M+1$ composite fermions have been divided into two partitions, $\{ z_1, z_2, \cdots z_{M+1}\}$ and $\{ w_1, w_2, \cdots w_{M}\}$, occupying states $\phi^{\rm CF}(\{z_j\})$ and $\chi^{\rm CF}(\{w_k\})$, and the last term represents correlations between composite fermions in different partitions. The symbol ${\cal A}$ indicates antisymmetrization with respect to exchange of any two particles. Let us define the largest exponent of $z_j$ in $\phi$ as $2Q_{\phi}$ and the largest exponent of $w_k$ in $\chi$ as $2Q_{\chi}$, which are analogous to the flux (measured in units of the flux quantum $\phi_0=hc/e$) in the spherical geometry for $\phi$ and $\chi$. Because the net flux must be the same for all particles, we have $2Q=2Q_{\phi}+M=2Q_{\chi}+M+1$ including the contribution from the cross factor. At $2Q=2N-3$, the flux value relevant for the 5/2 state, $\phi$ is the 1/3 state with a quasiparticle and $\chi$ is the $1/3$ state with a quasihole (Fig.~\ref{schematic}a). The addition of a CF in one partition thus induces, because of the structure of the wave function, a charge-neutral exciton across partitions, referred to as a ``topological" exciton.\cite{Hansson,Hansson09} It is to be distinguished from the ``ordinary" exciton (Fig.~\ref{schematic}b) that contains a quasiparticle-quasihole pair within one partition.\cite{topological}  Different occupations of the quasiparticle and the quasihole generate a basis of states for the UCF. The form of $\Psi^{\rm UCF}$ is naturally suggested by the CF theory\cite{Jain} 
within the bipartite representations of the paired Pfaffian wave function,\cite{Greiter} and has proved successful for the ordinary exciton.\cite{Sreejith} The conformal field theory construction of topological exciton by Hansson\cite{Hansson,Hansson09} is very similar, and may even be identical.  Wave functions of similar form were also motivated by Hermanns \cite{Hermanns} and by Milovanovi\' c and Jolic\oe ur\cite{Milovanovic} in a conformal field theory approach. 

\begin{figure}
\includegraphics[width=0.45\textwidth]{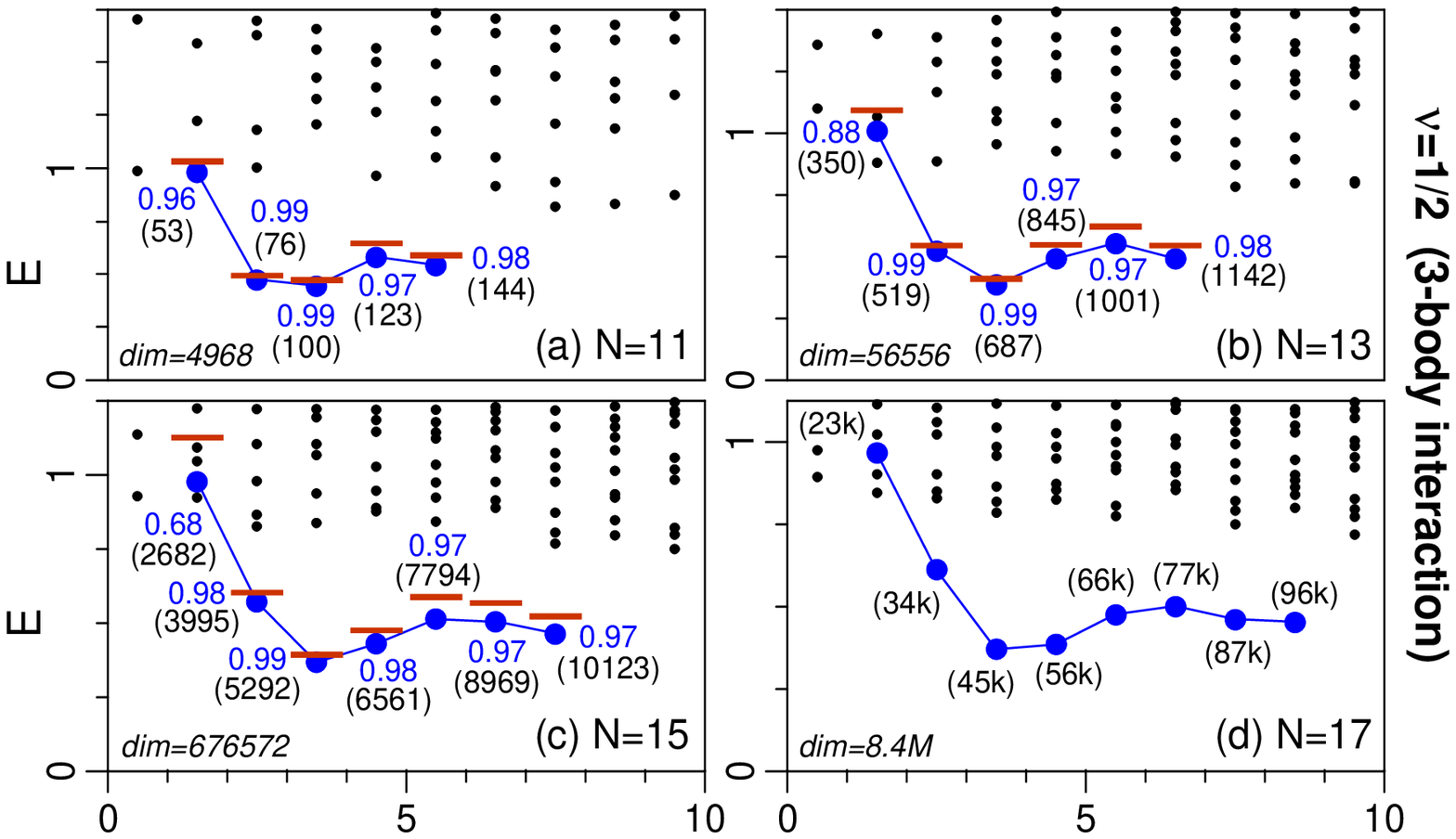}
\includegraphics[width=0.45\textwidth]{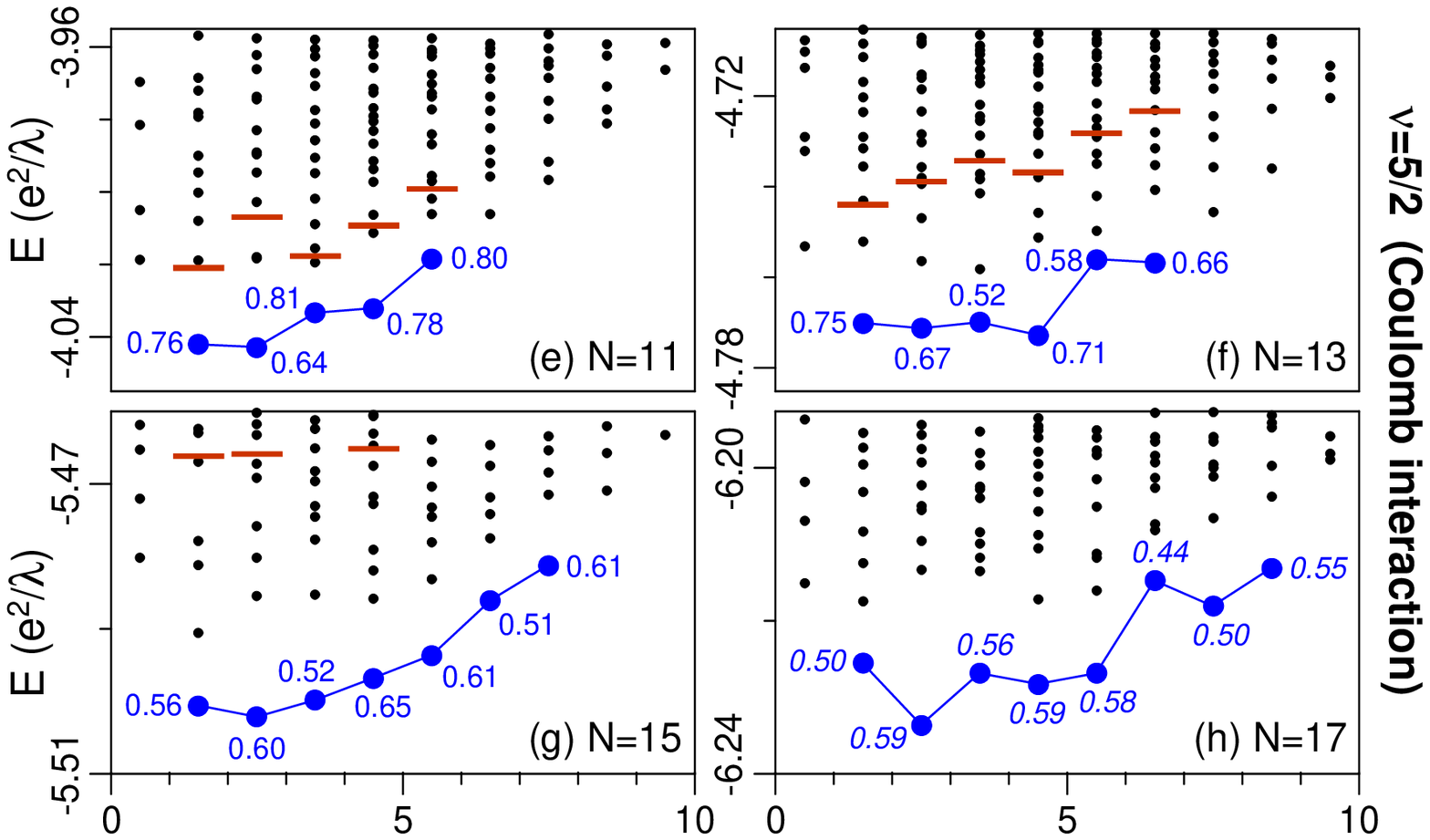}
\includegraphics[width=0.45\textwidth]{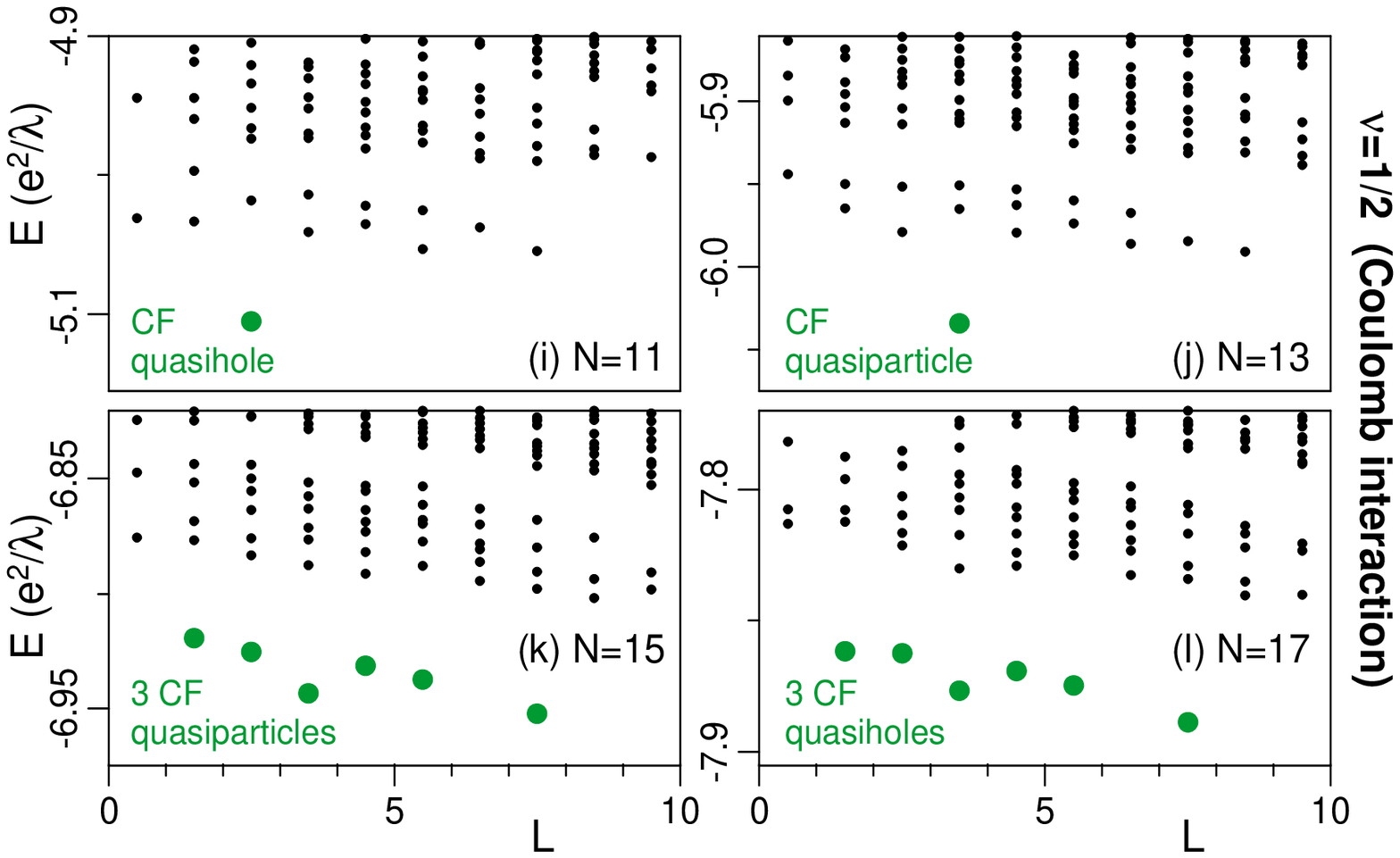}
\caption{(color online) 
Exact spectra (dots) for several particle numbers $N$ at total flux $2Q=2N-3$ for the 3-body interaction (a--d), for the Coulomb interaction in the 2nd LL (e--h), and for Coulomb interaction in the lowest LL (i--l). $L$ is the orbital angular momentum and $\lambda$ is the magnetic length. The fractional numbers near dots are overlaps with $\Psi^{\rm UCF}$ (which contains no adjustable parameters) in (a-c) and (e-g). The red dashes are the energies of $\Psi^{\rm UCF}$, not shown when they fall outside the frame (as in (g)). In (h) we show overlaps with the 3-body eigenstates; their Coulomb energies all fall outside the frame.  The total number of linearly independent states in each $L$ sector is shown in brackets in (a-d), and ``{\em dim}" indicates the dimension of the $L_z=1/2$ basis used in exact diagonalization.
\vspace*{-6mm}
}
\label{fig2}
\end{figure}

The ansatz $\Psi^{\rm UCF}$ signifies definite predictions, {\em with no free parameters}, for the quantum numbers of the low energy states, their wave functions, and their energies, through which the theory opens itself to rigorous tests against exact results known for finite systems. We show below that $\Psi^{\rm UCF}$ are exceedingly accurate for a model 3-body interaction $H_3$, and provide evidence that they are adiabatically connected to the solutions of the 2nd Landau level (LL) Coulomb Hamiltonian $H_{\rm C}$. This logic is analogous to the one used by Storni, Morf and Das Sarma\cite{Storni} to justify the Pfaffian ground state for even $N$. The calculations are performed in the spherical geometry in which the $N$ electrons move on the surface of the sphere under the influence of a radial magnetic field. The total flux through this spherical surface is $2Qhc/e$. The wave functions $\Psi^{\rm UCF}$ can be translated into the spherical geometry using standard methods. For the quasiparticle, Jain's wave function has been used.\cite{Jain,Jain2} The model 3-body interaction\cite{Greiter} is given by $H_3=\sum_{i<j<k}P^{(3)}_{ijk}(3Q-3)$, where $P^{(3)}_{ijk}(L)$ projects the state of the three particles $(i,j,k)$ into the subspace of total orbital angular momentum $L$; the interaction penalizes the smallest approach of three particles. We find it convenient to express the UCF wave function in each angular momentum sector as a linear combination of exact eigenstates, $\psi_{\rm exact}^{i}$, of either $H_3$ or $H_{\rm C}$ in the same sector:
$\Psi^{\mathrm{UCF}}=\sum_i c_{i}\psi_{\rm exact}^{i}$. By considering sufficiently many $N$-particle configurations, we obtain a system of linear equations which can be solved to obtain $c_i$. The energies and overlaps of $\Psi^{\rm UCF}$ are then evaluated straightforwardly.\cite{comment3}

\begin{figure}
\includegraphics[width=0.45\textwidth]{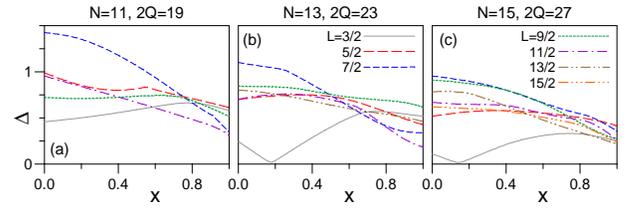}
\caption{(color online) 
Adiabatic evolution of the energy gap $\Delta(x)$ at each relevant value of $L$ (defined as the separation between the two lowest eigenvalues at that $L$) as the interaction is varied from 3-body to 2nd LL Coulomb. 
The evolution hamiltonian is $H(x)=(1-x)H_3/\varepsilon_{\rm 3}+xH_{\rm C}/\varepsilon_{\rm C}$ in which the energy scales $\varepsilon_3$ and $\varepsilon_{\rm C}$ are taken as the ordinary exciton energies at large wave vector. 
\vspace*{-6mm}
}
\label{fig3}
\end{figure}

The angular momentum of the CF-quasiparticle in $\phi^{\rm CF}$ is $(N+1)/4$ and the angular momentum of the CF-quasihole in $\chi^{\rm CF}$ is $(N-1)/4$, which gives the allowed angular momenta for their combination as $L=1/2$, $\cdots$, $N/2$. It turns out, remarkably, that the state at $L=1/2$ is exactly annihilated\cite{comment2} upon antisymmetrization, thus leaving the UCF states at $L=$ 3/2, 5/2, $\cdots$, $N/2$.  The annihilation of the state at the smallest $L$ is analogous to the annihilation of the $L=1$ CF exciton of the FQHE states at $\nu=n/(2n\pm 1)$, as noted by Dev and Jain.\cite{Dev92} 

The exact 3-body spectra (dots) are shown in the four upper panels (a)--(d) of Fig.~\ref{fig2} for several $N$. A low energy branch of states (blue dots) is seen to be well separated from the continuum, and with the exception of $L=3/2$, the angular momenta of these states match nicely with the predicted values. The 3-body energies of $\Psi^{\rm UCF}$ and their overlaps with the corresponding exact eigenstates are also shown. The exquisite agreement establishes the validity of $\Psi^{\rm UCF}$ for $H_3$,  with the exception of $L=3/2$ where $\Psi^{\rm UCF}$ better describes an excited state. 

\begin{figure}
\includegraphics[width=0.45\textwidth]{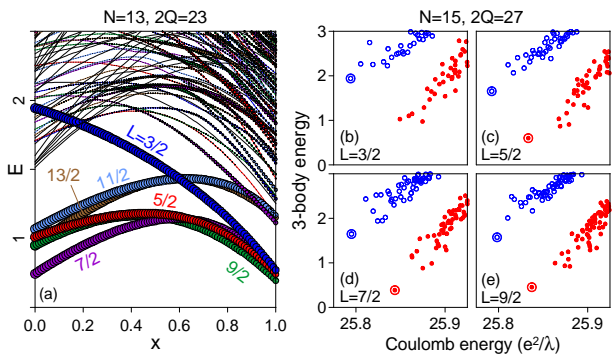}
\caption{(color online) Panel (a): Evolution of the low energy spectrum of $N=13$ electrons with the hamiltonian $H(x)$ of Fig.~\ref{fig3}. Both constituents $H_3$ and $H_{\rm C}$ are measured from the uniform ground state energy (interpolated to an odd $N$). Colored lines with dots have been used for the angular momenta of the UCF, and the dot diameters give the overlaps with the corresponding 3-body eigenstate. Except for $L=3/2$ (level crossing at $x\approx0.2$), the UCF states at $x=0$ appear adiabatically connected to the corresponding Coulomb states at $x=1$.
Panels (b--e): Blue open circles show $\langle \Psi_{\rm C}|H_3|\Psi_{\rm C}\rangle$  for each 2nd LL Coulomb eigenstate $\Psi_{\rm C}$ as a function of its eigenenergy $E_{\rm C}$, and the red dots show $\langle \Psi_3|H_{\rm C}|\Psi_3\rangle$ for each 3-body eigenstate $\Psi_3$ as a function of its eigenenergy $E_3$. The encircled red and blue dots indicate that the lowest 3-body state has the lowest $\left<H_{\rm C}\right>$ and {\em vice versa}, supporting adiabatic continuity; the red dots at $L=3/2$ provide an exception.  
\vspace*{-6mm}
}
\label{fig4}
\end{figure}

The exact spectra for the $H_{\rm C}$ are shown in panels (e)--(h) of Fig.~\ref{fig2}. These also contain a band of low-energy states (blue dots) at $L=$ 3/2, 5/2, $\cdots$, $N/2$, which is less well defined than the band for $H_3$, but the lowest state is well separated from the ``continuum" at each $L$ in this range. Unlike for $H_3$, the $L=3/2$ state does not belong in the continuum. The overlaps of the Coulomb eigenstates with $\Psi^{\rm UCF}$ are moderate; this is to be expected because 
the overlap of the exact 5/2 Coulomb ground state at even $N$ with the Pfaffian wave function are also of similar level, and there is no reason why $\Psi^{\rm UCF}$ should do better than the Pfaffian wave function.
We plot in Fig.~\ref{fig3} the evolution of the gap at each $L$ as we vary the interaction from $H_3$  to $H_{\rm C}$. The gap does not close for $L\geq 5/2$. There is an avoided level crossing at $L=3/2$, but, interestingly, the Coulomb state is seen to be connected to the 3-body {\em excited} state that has largest overlap with the UCF wave function.  These results suggest that the Coulomb eigenstates are adiabatically connected to $\Psi^{\rm UCF}$ for {\em all} relevant $L$.  Further evidence for adiabatic continuity is presented in Fig.~\ref{fig4}. We note that, as for the Pfaffian state, the agreement improves slightly (not shown) upon including finite thickness effects.

In contrast, the spectra for the {\em lowest} LL Coulomb state, shown in panels (i)--(l) of Fig.~\ref{fig2}, are consistent with a system of weakly interacting composite fermions experiencing an effective flux $2|Q^*|=2|Q-N+1|=1$. In the topmost partially filled $\Lambda$ level shell, we have: one CF hole with single particle angular momentum $l^*=5/2$  for $N=11$; one CF with $l^*=7/2$  for $N=13$; three CFs each with $l^*=7/2$ for $N=15$; and three CF holes each with $l^*=7/2$ for $N=17$.  The allowed $L$ values can be obtained from an elementary calculation, and match precisely those seen in the exact spectra. The similarity of the lowest bands for 15 and 17 particles is striking; there is no symmetry in the electron problem which implies this result, but it is explained rather naturally in the CF theory, where the two states are related by particle hole symmetry in the fourth $\Lambda$ level.  States with excitations across one $\Lambda$ level can also be identified in these spectra as forming a well defined second band. Clearly, the structure of the low energy states in the lowest LL is qualitatively distinct from that for $H_3$, implying a lack of adiabatic continuity with $H_3$.

The above description of the UCF gives natural insight into many properties of the 5/2 state:

{\em Odd-even parity effect:} The presence of the topological exciton at odd $N$ causes O(1) oscillations in energy as a function of $N$, as found by Lu, Das Sarma and Park,\cite{Hantao} with the energy difference between odd and even $N$ being equal to the energy of a topological exciton.

\begin{figure}
\begin{center}
\includegraphics[width=0.45\textwidth]{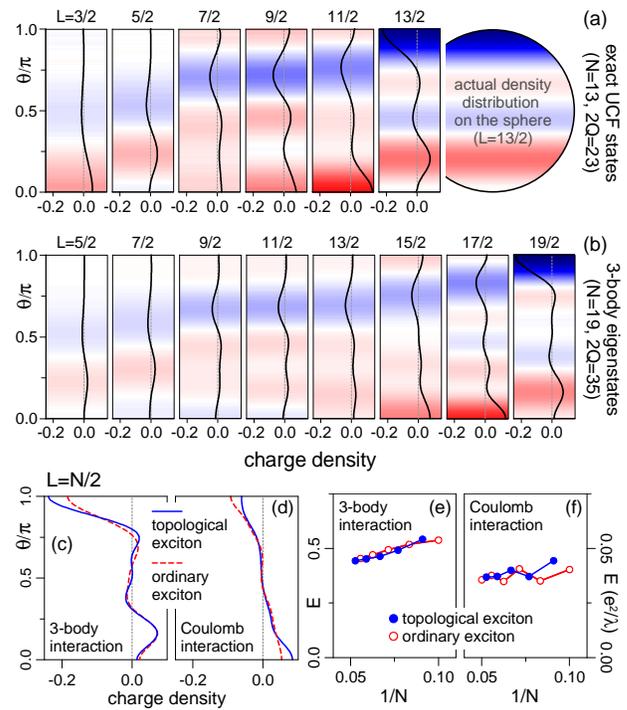}
\vspace*{-3mm}
\end{center}
\caption{\label{fig5} (Color online)
Density profiles of (a) the unpaired CF state $\Psi^{\rm UCF}$ with $N=13$, and (b) 3-body eigenstate ($N=19$) on sphere as a function of the polar angle $\theta$. The density, normalized to the filling factor, is measured relative to 1/2. The different curves correspond to different $L$ with $L_z=L$; $L=3/2$ is not shown in (b) because it falls into the continuum. The distance between the constituent quasiparticle and quasihole increases with $L$; they are both near the North Pole ($\theta=0$) for $L=3/2$ but at the opposite Poles for $L=N/2$. Panels (c) and (d) show comparisons of the density profiles of the topological exciton (with $N=19$) and ordinary excitons ($N=18$) at the largest $L=L_z=N/2$ for 3-body and 2nd LL Coulomb interactions; panels (e) and (f) show their energies as a function of $N$.  The dimension of the $L_z=1/2$ configuration space exceeds 107 million for $N=19$ at $2Q=35$, the largest so far for which $H_3$ has been diagonalized.
\vspace*{-6mm}
}
\end{figure}

{\em Ordinary vs. topological exciton:} A nontrivial consequence is that in the large wave vector limit (namely the large $L$ limit in the spherical geometry), the UCF and ordinary neutral exciton have the same energy because, in this limit, the constituent quasiparticle and quasiholes are independent and their energy does not depend on whether they reside in the same or different partitions. (The energy must be defined properly relative to the uniform ground state, which for odd $N$ is to be obtained by interpolation.)  For $H_3$ this has already been noted in exact diagonalization studies by M\"oller, W\'ojs and Cooper.\cite{Moller} Similar demonstration for $H_{\rm C}$ would be crucial; here, while finite size effects are stronger, but Fig.~\ref{fig5} convincingly demonstrates that even for $H_{\rm C}$ the two excitons at large $L$ converge with increasing $N$. At short distances (small $L$), on the other hand, the energies of the ordinary and topological excitons are {\em not} equal.  Further, the ordinary CF exciton is known to display a complex dispersion, possibly with several ``roton" minima, resulting from a complex interplay between the density profiles of the quasihole and the quasiparticle as a function of their separation.\cite{Dev92,Girvin,Scarola00,Pinczuk} Similar behavior can be expected for the topological exciton, the constituents of which also have complicated density profiles (Fig. \ref{fig5}), and indeed, the dispersions in Fig.~\ref{fig2} or previous studies\cite{Moller,Bonderson} do exhibit minima.  The minimum energy of the UCF will thus be lower than its large $L$ limit, as indicated by numerical studies.\cite{Bonderson} The location of the minimum depends on the form of the interaction, and will in general occur at different $L$ for $H_3$ and $H_{\rm C}$.\cite{Bonderson}  It is noted that light scattering, which does not alter $N$, will excite the {\em ordinary} exciton for states containing either even or odd $N$, which differ only by a single localized topological exciton.

{\em Fusion rules:} 
Using the standard terminology of the Ising conformal field theory,\cite{NayakRMP} we identify the UCF by $\psi$ and the vortex excitation (a quasiparticle or a quasihole in one partition) by $\sigma$.  The relation $\sigma\times \sigma=1+\psi$ indicates that a quasihole and a quasiparticle can be combined to produce two kinds of excitations, by placing them in the same or different partitions. The former, labeled ``1,"  produces an ordinary exciton all of whose quantum numbers are zero, whereas the latter, labeled ``$\psi$," produces a UCF.  The relation $\sigma\times \psi=\sigma$ captures the reaction in which the addition of a quasiparticle or a quasihole to the UCF state annihilates half of the UCF to leave a single quasiparticle or quasihole. Finally, $\psi\times \psi=1$ encapsulates the fact that two UCFs are equivalent to two ordinary excitons. In all cases above, we have considered the lowest energy outcomes only.  The fusion relations in the presence of several quasiparticles or quasiholes can similarly be derived.

{\em Zero modes:} M\"oller, W\'ojs and Cooper\cite{Moller} have found that the energy remains unchanged, modulo finite size uncertainties, when a UCF is added to a system that already contains some quasiparticles or quasiholes. To see how this can be understood, take the example of a state with $2n$ quasiholes, $n$ in each partition. Adding a UCF produces an ``imbalanced" system with $n+1$ and $n-1$ quasiholes in the two partitions, because the quasiparticle of the topological exciton annihilates one of the quasiholes. Provided the quasiholes are far apart, the energy both before and after is simply $2n$ times the self energy of an isolated quasihole. The addition of a UCF thus costs zero energy. 

One may question if these conclusions, which rely on the validity of $\Psi^{\rm UCF}$, apply to the solutions of the 3-body or Coulomb interaction. In this context, it is important to note that the analysis leading to these conclusions does not depend on the details of the wave functions but only on the structure of the theory for the quasiparticles and quasiholes, and the existence of a topological exciton associated with the UCF. Adiabatic continuity suggests that the above mentioned implications carry over to $H_3$ and, likely, also to $H_{\rm C}$, and should be robust to corrections arising from finite width and Landau level mixing. 

We are grateful to Duncan Haldane, Maria Hermanns, Perla Kacman, Gunnar M\"oller, and especially Hans Hansson, for insightful discussions. We acknowledge financial support from the DOE under grant no. DE-SC0005042 and the Polish MNiSW Grant No. N202179538, and thank the Penn State High-Performance Computing Cluster for computer time. 

\newcommand{\PRL}{Phys.\ Rev.\ Lett.}
\newcommand{\PRB}{Phys.\ Rev.\ B}
\newcommand{\PRD}{Phys.\ Rev.\ D}
\newcommand{\NPB}{Nucl.\ Phys.\ B}


\begin{thebibliography}{99}
\bibitem{comment1} The unpaired composite fermion has been referred to as a ``neutral fermion" in the literature.\cite{Greiter,Moller,Bonderson}
\bibitem{topological} The terminology notwithstanding, we should remember that, because composite fermions themselves are topological, both excitons are fundamentally topological.
\bibitem{comment3} For 17 particles, the dimension of the $L_Z=1/2$ basis, shown on Fig.~\ref{fig2}, is too large for this method.
\bibitem{comment2} We do not have an analytic proof for this, but have found it to be correct to machine accuracy for $N\leq 15$.
\end{thebibliography}
\end{document}